\documentclass{iopart}
\usepackage{graphicx}

\begin{document}

\letter{Intrasubband plasmons in weakly disordered array of quantum wires}
\jl{3}
\author{Y.V. Bludov}
\address{Usikov Institute for Radiophysics and Electronics, NAS of Ukraine, 12 Acad. Proscura St., Kharkov, 61085, Ukraine Email: bludov@ire.kharkov.ua}

\begin{abstract}
The paper deals with the theoretical investigation of plasmons in
weakly disordered array of quantum wires, consisting of finite
number of quantum wires, arranged at an equal distance from each
other. The array of quantum wires is characterized by the fact
that the density of electrons of one "defect" quantum wire was
different from that of other quantum wires. At the same time it is
assumed that "defect" quantum wire can be arranged at an arbitrary
position in array. It is shown that the amount of plasmon modes in
weakly disordered array of quantum wires is equal to the number of
quantum wires in array. The existence of the local plasmon mode,
whose properties differ from those of usual modes, is found. We
point out that the local plasmon mode spectrum is slightly
sensitive to the position of "defect" quantum wire in the array.
At the same time the spectrum of usual plasmon modes is shown to
be very sensitive to the position of "defect" quantum wire.
\end{abstract}
\pacs{7320M, 7867L}
\maketitle

Collective charge-density excitations, or plasmons in quantum
wires (QW) are the objects of great physicist's interest. Earlier
plasmons in QW were investigated both theoretically
\cite{dsfirst,dsb,gold,ldj,hw} and experimentally
\cite{han,dem,goni}. In that papers it was shown that plasmons in
QW possess some new unusual dispersion properties. Firstly, the
plasmon spectrum depends strongly on the width of QW. Secondly, 1D
plasmons are free from the Landau damping \cite{dsb,hw} in the
whole range of wavevectors.

From the point of view of practical application so-called weakly
disordered arrays of low-dimensional systems are the objects of
interest. Recently the plasmons in weakly disordered
superlattices, formed of an finite number of equally spaced
two-dimensional electron systems (2DES), were theoretically
investigated \cite{gvozd,js,sy,ya}. The weakly disordered
superlattice is characterized by the fact that all 2DES possess
equal density of electrons except one (''defect'') 2DES, which
density of electrons differs from that of other 2DES. It was found
that the plasmon spectrum of such a superlattice contains the
local plasmon mode, whose properties differ from those of other
plasmon modes. Notice that practically all flux of electromagnetic
energy of plasmons, which correspond to the local mode, are
concentrated in the vicinity of ''defect'' 2DES. At the same time
paper \cite{ya} denoted the opportunity of using the plasmon
spectrum particularities to determine the parameters of defects in
superlattice.

This paper deals with the theoretical investigation of plasmons in
weakly disordered array of QW, consisting of a finite number $M$
of QW, arranged at planes $z=ld$ ($l=0,...,M-1$ is the number of
QW, $d$ is the distance between adjacent QW). We suppose that all
QW possess equal 1D density of electrons $N$ except one ''defect''
QW whose density of electrons is equal to $N_{\rm{d}}$. So, the
density of electrons in $l$-th QW can be expressed as
$N_l=(N_{\rm{d}}-N)\delta_{pl}+N$. Here $p$ is the number of
''defect'' QW arranged at the plane $z=pd$, $\delta_{pl}$ is the
Cronecker delta. QW are considered to be placed into the uniform
dielectric medium with dielectric constant $\varepsilon$. We
consider the movement of electrons to be free in $x$-direction and
is considerably confined in directions $y$ and $z$. At the same
time we suppose that the width of all QWs is equal to $a$ in
$y$-direction and is equal to zero in $z$-direction. By other
words, each QW can be represented as a square quantum well with
infinite barriers at $y=-a/2$ and $y=a/2$ and zero thickness in
$z$-direction. At the same time we take into account only the
lowest subband in each QW. In that case the single-particle wave
function for the electron can be written in the form:
\begin{equation}
\psi_{k_x,l}(x,y,z)=|k_x,l\rangle= e^{\rmi
k_xx}\varphi(y)\left[\delta(z-ld)\right]^{1/2}, \label{wf}
\end{equation}
where $\varphi(y)=\displaystyle{\sqrt{\frac{2}{a}}\cos{\frac{\pi
y}{a}}}$, $k_x$ is the one-dimensional wave vector, describing the
motion in $x$-direction. In that case the single-particle energy
can be represented by such an expression: $$
E_{k_x,l}=E_0+\frac{\hbar^2k_x^2}{2m^*}.$$ Here $E_0$ is the
energy of subband bottom (for simplicity we may put $E_0=0$),
$m^*$ is the effective mass of the electron.

To obtain the collective excitations spectrum we start with a
standard linear-response theory in an random phase approximation.
To obtain the collective excitations spectrum we consider $\delta
n(\bf{r})$ which is the deviation of the electron density from its
equilibrium value. After using the standard linear-response theory
and the random phase approximation, $\delta n(\bf{r})$ can be
related to the perturbation as
\begin{equation}
\delta
n(x,y,z)=\sum_{\alpha,\alpha'}\frac{f_{\alpha'}-f_{\alpha}}{E_{\alpha'}-E_{\alpha}+\hbar\omega}V_{\alpha\alpha'}\psi_{\alpha'}^*\psi_{\alpha},
\label{nxyz}
\end{equation}
where $\alpha=(k_x,l)$ is a composite index which is defined by
(\ref{wf}) and $f_{\alpha}$ is the Fermi distribution function,
$V_{\alpha,\alpha'}=\langle\alpha|V|\alpha'\rangle$ are the matrix
elements of the perturbing potential $V=V^{ex}+V^H$, $V^{ex}$ and
$V^H$ are the external and Hartree potentials, correspondingly.

For our system equation (\ref{nxyz}) can be rewritten in the form
\begin{equation}
\delta n(q_x,y,z)=\sum_{l'}\Pi^{l'}V_{l'}\phi^2(y)\delta(z-l'd),
\label{nq}
\end{equation}
where $
\displaystyle{\Pi^{l'}=\frac{1}{\pi}\int_{-\infty}^{\infty}dk_x\frac{f_{k_x+q_x,l'}-f_{k_x,l'}}{E_{k_x+q_x,l'}-E_{k_x,l'}+\hbar\omega}}
$ is the noninteracting 1D polarizability ("bare bubble")
function, $V_{l'}\equiv \langle k_x,l'|V|k_x+q_x,l'\rangle$. At
zero temperature function $\Pi^l$ can be written as
\begin{equation}
\Pi^l=\frac{m^*}{q_x\pi\hbar^2}\ln{\frac{\omega^2-\displaystyle{\left(\frac{\hbar
q_xk^l_F}{m^*}-\frac{\hbar
q_x^2}{2m^*}\right)^2}}{\omega^2-\displaystyle{\left(\frac{\hbar
q_xk^l_F}{m^*}+\frac{\hbar q_x^2}{2m^*}\right)^2}}}.
\end{equation}
Here $\displaystyle{k^l_F=\frac{\pi N_l}{2}}$ is the Fermi
wavenumber in $l$-th QW. In the long-wavelength limit (where $q_x
\to 0$) function $\Pi^l$ can be written as
$\displaystyle{\Pi^l=\frac{N_l}{m^*}\frac{q_x^2}{\omega^2}}$.

Notice, that the Hartree potential can be expressed through the
perturbation \cite{dsb} as
\begin{equation}
V^H(\bf{r})=\int \rmd\textbf{r}'
\frac{\it{e^2}}{\varepsilon|\textbf{r}-\textbf{r}'|}\delta {\it
n}(\textbf{r}'). \label{hp}
\end{equation}
Using equations (\ref{nq}) and (\ref{hp}) we get such an
expression for matrix element $V^H_l=\langle
k_x,l|V^H(x,y,z)|k_x+q_x,l\rangle$:
\begin{equation}
V^H_l=\sum_{l'}\Pi^{l'}V_{l'}U_{l,l'}, \label{ursex}
\end{equation}
where $$U_{l,l'}=\frac{8e^2}{\varepsilon a^2}\int_{-a/2}^{a/2}\rmd
y'\int_{-a/2}^{a/2}\rmd y
K_0(q_x[(y-y')^2+(l-l')^2d^2]^{1/2})\cos^2{\left(\frac{\pi
y}{a}\right)}\cos^2{\left(\frac{\pi y'}{a}\right)},$$ $K_0(x)$ is
the zeroth-order modified Bessel function of the second kind.
Collective excitations of QW array exist when equation
(\ref{ursex}) has nonzero solution $V^H$ in the case where the
external perturbation $V^{ex}=0$. Hence, the intersubband plasmon
dispersion relation has the form
\begin{equation}
\det|\delta_{l,l'}-\Pi^{l'}U_{l,l'}|=0. \label{dre}
\end{equation}
It should be noticed that when $M=2$ the dispersion relation
(\ref{dre}) coincides with the dispersion relation for plasmons in
double-layer QW system, obtained in \cite{dsb}.

\begin{figure}[t!]
  \centering %
  {\scalebox{1}[1]{\includegraphics[16,676][386,816]{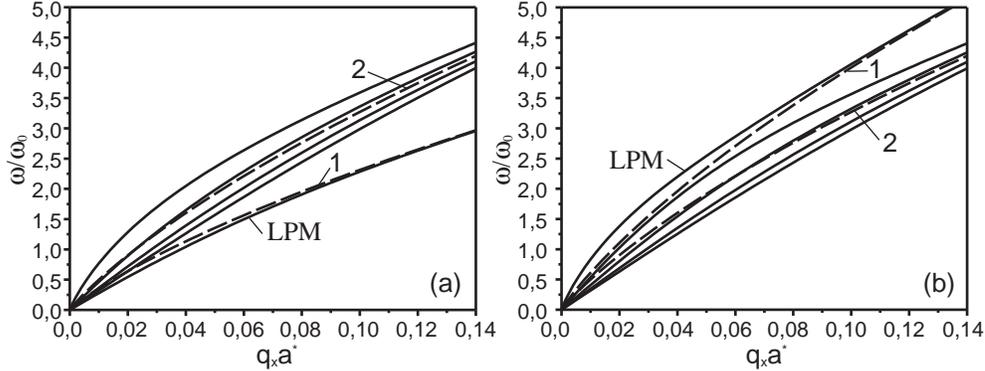}}}%
  \caption{Dispersion curves of plasmons  (solid curves) in weakly
  disordered array of QW with parameters $M=5$, $d=15a^*$, $a=20a^*$, $p=0$
  for two values of the density of electrons of "defect" QW: (a) $N_{\rm{d}}/N=0,5$, (b) $N_{\rm{d}}/N=1,5$.
  For comparison the dispersion curves for the plasmons in single QW
  with electron density $N_{\rm{d}}$ and $N$ are depicted by dashed curves
  1 and 2, correspondingly.}
  \label{fg:drel}
\end{figure}

\Fref{fg:drel} shows the intrasubband plasmon spectrum (solid
lines) in weakly disordered array of QW in the case where $p=0$.
The $y$-axis gives the dimensionless frequency $\omega/\omega_0$
($\omega_0^2=2Ne^2/\varepsilon m^*a^2$ is the plasma frequency),
and the $x$-axis gives the dimensionless wavevector $q_xa^*$
($a^*=\varepsilon\hbar^2/m^*e^2$ is the effective Bohr radius). As
the model of QW we use heterostructure GaAs with the effective
mass of electrons $m^*=0,067m_0$ ($m_0$ is the mass of free
electron) and with the dielectric constant $\varepsilon=12,7$.

As seen from \fref{fg:drel}, the intrasubband plasmon spectrum in
finite array of QW contains $M$ modes. So, the number of modes in
the spectrum is equal to the number of QW in the array. Notice,
that with the increase of wavenumber $q_x$ the plasmon frequency
$\omega$ also increases. At the same time the propagation of
plasmons in weakly disordered array of QW is characterized by the
presence of local plasmon mode (LPM). In the case where the
density of electrons in ''defect'' QW is less then the density of
electrons in other QW ($N_{\rm{d}}<N$), the LPM lies in the
lower-frequency region in comparison with the usual plasmon modes
(\fref{fg:drel}a). Correspondingly, if $N_{\rm{d}}>N$, the LPM
lies in the higher-frequency region in comparison with the usual
ones (\fref{fg:drel}b). It should be emphasized that at the limit
$q_xd \to \infty$, when the Coulomb interaction between electrons
in adjacent layers is neglible, the LPM dispersion curve is close
to the dispersion curve for the plasmons in single QW with the
density of electrons $N_{\rm{d}}$. At the same time the dispersion
curves for usual plasmon modes at the limit $q_xd \to \infty$ are
gradually draw together and are close to the dispersion curve for
plasmon in the single QW with the density of electrons $N$.

\begin{figure}[t!]
  \centering %
  {\scalebox{1}[1]{\includegraphics[96,485][465,764]{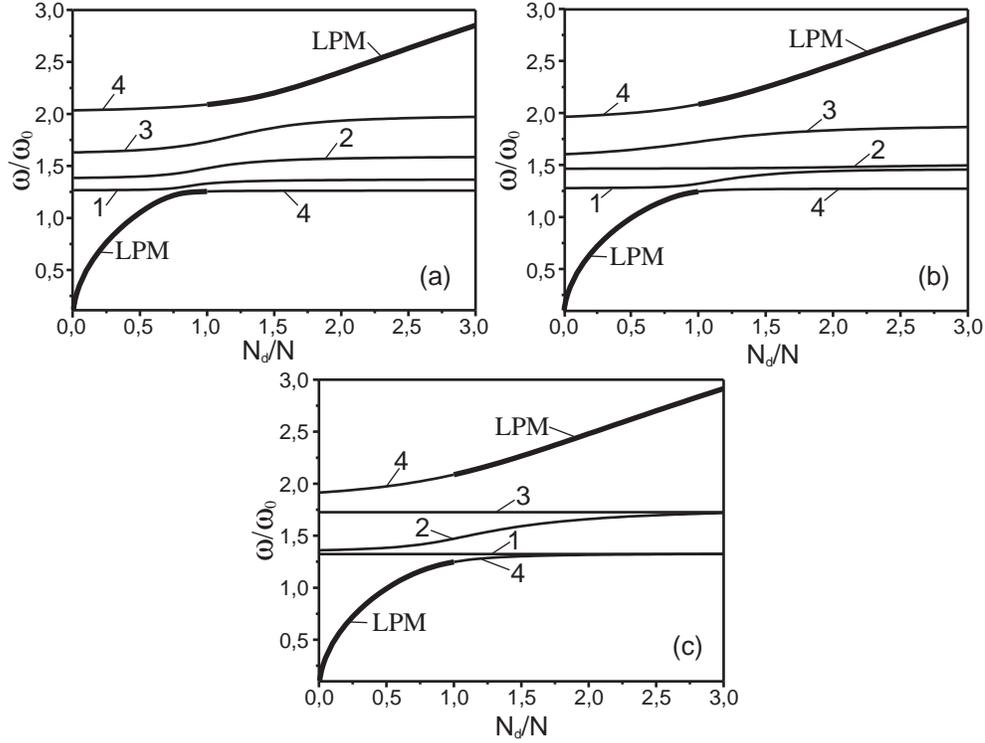}}}%
  \caption{Dependence of plasmon frequency upon the ratio $N_{\rm{d}}/N$
  when $q_xa^*=0,04$ and
  for three cases of "defect" QW position in the array: (a) $p=0$, (b) $p=1$, (c) $p=2$.}
  \label{fg:fiq}
\end{figure}

Now we consider the dependence of plasmon spectrum upon the value
of 1D electron density in ''defect'' QW. \Fref{fg:fiq} presents
the dependence of plasmon frequency upon the ratio $N_{\rm{d}}/N$
in the case of fixed value of wavevector $q_x$ and for different
positions of the ''defect'' QW in the array. As seen from
comparison of \fref{fg:fiq}a,b,c the LPM spectrum depends weakly
upon the position of ''defect'' QW in array. However, the spectrum
of usual plasmon modes is more sensitive to the position of
''defect'' QW in the array. Notice, that the frequency of LPM
increases when the value of ratio $N_{\rm{d}}/N$ is increased. At
the same time the usual plasmon modes spectrum is characterized by
such a features. As $p=0$ (\fref{fg:fiq}a) when the value of ratio
$N_{\rm{d}}/N$ is increased, the frequency of usual plasmon modes
also increases. However when $p=1$ (\fref{fg:fiq}b) the frequency
of one of the usual plasmon modes (curve 2) does not practically
depend upon the value of ratio $N_{\rm{d}}/N$. In the case where
$p=2$ (\fref{fg:fiq}c) there are already two plasmon modes (curves
1 and 3) which possess such a particularity.

In summary, we calculated the plasmon spectrum of finite weakly
disordered array of QW, which contains one "defect" QW. It is
shown that amount of plasmon modes in the spectrum is equal to the
amount of QW in array. It is found that the LPM, whose properties
differ from those of other modes, exists in the plasmon spectrum.
We point out that the LPM spectrum is slightly sensitive to the
position of "defect" QW in array. At the same time position of
"defect" QW exerts influence on the spectrum of usual plasmon
modes. It is shown that under certain conditions the existence of
plasmon modes, which spectrum does not depend upon density of
electrons of "defect" QW, is possible. Notice, that the
above-mentioned features of plasmon spectra can be used for
diagnostics of defects in QW structures.

\end{document}